\documentclass[preprint,NumberedRefs]{JASA}

\let\JASAsavedthinmuskip\thinmuskip
\let\JASAsavedtmspace\tmspace

\let\linenumbers\relax
\AtBeginDocument{\nolinenumbers}

\begin{document}

\title[Multiple spontaneous otoacoustic emissions as a single standing wave]{Multiple spontaneous otoacoustic emissions as a single standing wave unstable to modulation}


\author{Yasuki Murakami}

\affiliation{Faculty of Design, Kyushu University, 4-9-1 Shiobaru, Minami-ku, Fukuoka 815-8540, Japan}

\email{murakami@design.kyushu-u.ac.jp}

\begin{abstract}
Spontaneous otoacoustic emissions (SOAEs) have been interpreted either as
cochlear standing waves or as local oscillations of hair cells. Multiple
SOAE lines test both views, because each view must explain how the lines
coexist. In the standing-wave view, each line is a separate resonance. In
the intermodulation view, some lines are distortion products of others.
Both explanations assume several independent oscillations. Independent
oscillations, however, compete through the saturating cochlear amplifier,
and the dominant oscillation suppresses the others. The saturation has
another role besides suppression. The saturation also induces an
oscillation of the gain at twice the frequency of the oscillation. In
lasers, this oscillation of the gain, called the \textit{pump}, creates sidebands.
This paper examines whether a cochlear standing wave can pump the gain by
itself. A two-dimensional cochlear model with a step in the gain produced a
single standing wave. The pump destabilized a pair of lines near the
neighboring resonances. The other lines were phase-locked intermodulation
products of the main line and the pair. A group of equally spaced SOAEs can
thus arise from one self-sustained oscillation. The spacing follows the
resonances of the standing wave, and the two explanations can hold together.
\end{abstract}

\maketitle

\section{Introduction}
\label{sec:intro}

Spontaneous otoacoustic emissions (SOAEs) are narrow-band sounds that the ear
emits without stimulation~\cite{Kemp1978}, and the origin of SOAEs is not
settled~\cite{Wit2026}. In one view, mammalian SOAEs are amplitude-stabilized
cochlear standing waves~\cite{Shera2003}. Waves reflected within the cochlea
and at the stapes build up at the frequencies where the round-trip phase is a
multiple of $2\pi$ and the round-trip gain reaches unity. The saturation of
the cochlear amplifier fixes their amplitude~\cite{ZweigShera1995,Talmadge1998}.
In the other view, SOAEs come from critical (Hopf) oscillators of the hair
cells~\cite{Camalet2000,Hudspeth2010}. Coupled arrays of such oscillators
reproduce the statistics of human SOAEs~\cite{Fruth2014}. Human SOAEs have
also been modeled as the synchronized clusters of a chain of coupled van der
Pol oscillators~\cite{WitVanDijk2012}. Spectra with several lines can help
to decide, because each view must explain how the lines coexist.

Multiple SOAEs in one ear have been explained in two ways, as separate
resonances or as intermodulation products. In the standing-wave view, each
SOAE is a separate resonance. Neighboring SOAEs are then spaced by about the
inverse of the round-trip delay~\cite{Shera2003,Shera2022}. The same delay sets the period
of the fine structure of stimulus-frequency OAEs (SFOAEs) and of the
microstructure of hearing thresholds~\cite{SheraGuinan2003,Dewey2025}. In
the other view, some SOAEs are intermodulation products of others. In triplets of
equally spaced SOAEs, the phases of the three peaks lock as those of a cubic
distortion product~\cite{vanDijkWit1998}. Such triplets are the exception,
since the ratios of the frequency differences of consecutive SOAEs generally
do not cluster at unity, the value expected for
intermodulation~\cite{Bergevin2025}. The two explanations thus differ in
whether every SOAE is an independent oscillation.

Both explanations assume several independent oscillations, and independent
oscillations tend not to coexist. Independent oscillations share the
saturating cochlear amplifier, and a strong component suppresses the others,
as in two-tone suppression~\cite{Rhode2007}. In cochlear models, the
dominant oscillation suppresses the other modes~\cite{Ku2009}. The standing-wave view must therefore explain why
several resonances oscillate at once. The intermodulation view must likewise
explain why two or more primary oscillations survive. Locked phases alone do
not reveal the origin of a line, because coupled independent oscillators
also lock~\cite{Gelfand2010,WitVanDijk2012}. In cochlear models, the lines
other than the dominant one have been attributed to independent unstable
modes or to their distortion products~\cite{Ku2009,Meaud2018}. If a single
standing wave generated the other lines, the lines would coexist without
competing. The spacing would then follow the resonances, while the other
lines would be intermodulation products. The two views would hold together.

Analyses of cochlear models have treated the saturation only as an average
drop of the gain. The eigenvalues of a model at rest show which modes start
to grow~\cite{Elliott2007,Ku2008}. Once an oscillation saturates the
amplifier, the saturation has been included as a gain averaged over a
cycle~\cite{Ku2009}. The averaged gain is lower than the gain at rest. A
lower gain can only suppress the other lines, not create new lines.

The saturation, however, also induces an oscillation of the gain, and the
oscillating gain can create sidebands. An oscillation saturates the gain at each peak, twice
in each cycle. The gain therefore oscillates at twice the frequency of the
oscillation. This oscillation of the gain is called the \textit{pump} here.
In lasers, an oscillating gain creates a pair of lines on both sides of the
main line~\cite{Lamb1964,Agrawal1988}. There, the gain responds to the
intensity averaged over the optical cycle, and it oscillates at the beat
frequency of the main line and a sideband. In the cochlear amplifier, the
gain follows the displacement itself, and it oscillates at twice the
frequency even for a single line. Both oscillating gains couple the two
sidebands by four-wave mixing.
In ears, sidebands of an SOAE have been produced by an external
low-frequency tone~\cite{BianWatts2008}.
A cochlear standing wave may thus pump the gain by itself and create lines
near the neighboring resonances.

This paper shows that a group of equally spaced SOAE lines can arise from a
single cochlear standing wave that the pump renders unstable to modulation.
The mechanism concerns such a group from one standing wave, not SOAE
spectra in general (Sec.~\ref{sec:two}). The model is a
two-dimensional (2D) cochlea with a lumped middle ear~\cite{LiuNeely2010}.
Unlike distributed roughness~\cite{ZweigShera1995,Bowling2019}, a step in
the gain places the reflection at a single known
position~\cite{Nobili2003}, so the resonances are known. The eigenvalues of
the model about the standing wave, computed with the pump, show which lines
the pump destabilizes. The time-domain solution confirms the eigenvalues.

\section{Model}
\label{sec:model}

\subsection{Cochlear fluid}
\label{sec:fluid}

\begin{figure}[t]
\includegraphics[width=\linewidth]{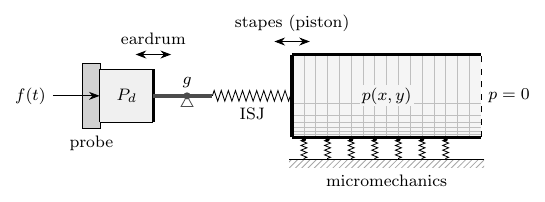}
\caption{\label{fig:model}The model. A probe drives the air in the ear
canal ($P_d$). The eardrum drives the cochlear fluid through the
malleus--incus lever ($g$), the incudo-stapedial joint (ISJ), and the
stapes. The stapes moves the basal face of the cochlea as a piston. The thin
lines indicate the discretization of the fluid in $x$ and $y$. The same
micromechanics is attached at each grid point of the BM, and $p = 0$ at the
helicotrema.}
\end{figure}

Figure~\ref{fig:model} shows the model, which consists of a probe in the ear canal, the middle ear, the cochlear fluid, and the micromechanics of the cochlear partition. Most cochlear models are one-dimensional transmission lines with a lumped middle ear~\cite{Talmadge1998,Verhulst2012}. Fluid models in 2D have also been integrated in the time domain~\cite{AllenSondhi1979}, and SOAEs and SFOAEs have been computed as the ear-canal pressure~\cite{Meaud2018,Vencovsky2020}. 

Here the cochlea is unrolled into a 2D box. The coordinate $x$ runs along the partition from the base ($x = 0$) to the helicotrema ($x = L$), and $y$ is perpendicular to the partition at $y = 0$. The two scalae, each of height $H$, are symmetric about the partition. Only $0 \le y \le H$ is therefore modeled, and $p$ is the pressure difference between the scalae, twice the pressure in the scala vestibuli. The fluid is ideal and incompressible, so that $p$ satisfies the Laplace equation,
\begin{equation}
\frac{\partial^2 p}{\partial x^2} + \frac{\partial^2 p}{\partial y^2} = 0.
\label{eq:laplace}
\end{equation}
The pressure couples to the displacement $u_b(x,t)$ of the basilar membrane (BM) at $y = 0$, positive toward the scala tympani ($-y$), the wall at $y = H$ is rigid, the stapes, with displacement $x_s$, moves the whole face $x = 0$ as a piston, and the pressure vanishes at the helicotrema ($x = L$). The boundary conditions are thus
\begin{align}
\frac{\partial p}{\partial y} &= 2\rho\,\ddot{u}_b \quad \text{at } y = 0, \label{eq:bc_bm}\\
\frac{\partial p}{\partial y} &= 0 \quad \text{at } y = H, \label{eq:bc_wall}\\
\frac{\partial p}{\partial x} &= -2\rho\,\ddot{x}_s \quad \text{at } x = 0, \label{eq:bc_stapes}\\
p &= 0 \quad \text{at } x = L, \label{eq:bc_helico}
\end{align}
where the dots denote time derivatives. In this study, $L = 35$~mm, $H = 7$~mm, and $\rho = 1000$~kg/m$^3$, as in Ref.~\cite{Sisto2019}.

\subsection{Viscous damping of the BM}
\label{sec:viscous}

An ideal fluid carries arbitrarily short waves along the BM. Such waves destabilize active models. The shortest waves of the model, and hence the results of the model, then depend on the grid. In the perilymph, viscosity damps such waves~\cite{Nobili2003}. The damping force on the BM grows with the wavenumber~\cite{Sisto2021}. Following that approach, the pressure is computed for the ideal fluid, and the viscosity enters only as a damping force on the BM.

A damping that grows with the wavenumber is not a local force. Such a damping was earlier written with the local wavenumber of a single traveling wave~\cite{Sisto2021}. In the time domain, the BM carries waves of many wavenumbers at once. The damping is therefore applied to the cosine modes of the BM velocity,
\begin{equation}
\dot{u}_b(x,t) = \sum_{q=0}^{\infty} \hat{v}_q(t) \cos k_q x,
\qquad k_q = \left(q + \tfrac{1}{2}\right)\frac{\pi}{L},
\label{eq:modes}
\end{equation}
where $k_q$ are the wavenumbers that satisfy the conditions at the base and the helicotrema~\cite{Murakami2021}. Each mode receives its own damping force, and $F_v$ is their sum,
\begin{equation}
F_v(x,t) = \sum_{q=0}^{\infty} 8\mu k_q \coth(k_q H)\, \hat{v}_q(t) \cos k_q x,
\label{eq:Fv}
\end{equation}
where $\mu$ is the viscosity. The coefficient equates the power of $F_v$ to the viscous dissipation of the ideal flow in both scalae~\cite{Lamb1932}. This dissipation is that of the irrotational flow in the bulk of the fluid, not the shear in the rotational viscous layer on the BM~\cite{Stokes1851}. The bulk dissipation dominates for the short waves that the damping must remove~\cite{Sisto2021}. The treatment holds while the viscous layer is thin compared with the wavelength, so that the flow outside the layer stays irrotational. The damping removes the short waves while hardly affecting the long ones. For $\mu = 0.7$~mPa\,s, the damping coefficient is 59~N\,s/m$^3$ for a wavelength of 0.6~mm and 350~N\,s/m$^3$ for 0.1~mm, against the BM damping of model 2 of Ref.~\cite{Sisto2019}, 880~N\,s/m$^3$, at $x = 10$~mm.

\subsection{Cochlear micromechanics}
\label{sec:micromechanics}

The partition at each point is the two-degree-of-freedom system of model 2 of Ref.~\cite{Sisto2019}. The two masses of the partition are the BM, with mass $m_1$ and displacement $u_b$, and the tectorial membrane (TM), with mass $m_2$ and displacement $u_t$, both positive in the same direction as $u_b$. Their equations of motion are
\begin{align}
m_1 \ddot{u}_b &= p - F_b - F_v, \label{eq:eom_b}\\
m_2 \ddot{u}_t &= -F_t, \label{eq:eom_t}
\end{align}
where $F_v$ is the viscous damping of Eq.~(\ref{eq:Fv}). The forces of the springs, the dampers, and the active element on the BM and the TM are
\begin{align}
F_b &= c_1 \dot{u}_b + k_1 u_b + c_3 \dot{u}_c + k_3 u_c
       - \gamma\,(c_4 \dot{u}_c + k_4 u_c), \label{eq:Fb}\\
F_t &= c_2 \dot{u}_t + k_2 u_t - c_3 \dot{u}_c - k_3 u_c
       + \gamma\,(c_4 \dot{u}_c + k_4 u_c), \label{eq:Ft}
\end{align}
where $u_c = u_b - u_t$. The masses, stiffnesses $k_i$, and damping coefficients $c_i$ are those of model 2 of Ref.~\cite{Sisto2019}. The active element is driven by $u_c$, and its force, the active force, is $\gamma\,(c_4 \dot{u}_c + k_4 u_c)$. The active force acts with opposite signs on the BM and the TM. The active force is limited by a cubic term,
\begin{equation}
\gamma = \gamma_0(x) \left( 1 - \frac{u_c^2}{u_{\mathrm{sat}}^2} \right),
\label{eq:gamma}
\end{equation}
where $\gamma_0(x)$ is the gain at low levels and $u_{\mathrm{sat}}$ is the saturation displacement, at which the gain falls to zero. This saturation is the only nonlinearity of the model. The active force thus contains the negative damping of a van der Pol oscillator, which decreases as $u_c$ approaches $u_{\mathrm{sat}}$. The active force also contains a corresponding cubic stiffness. Cubic saturation has been used in models of SOAEs and distortion products~\cite{Talmadge1998,Ku2009}.

The cubic term has no linear range and renders the active element passive for $|u_c| > u_{\mathrm{sat}}$. This cubic form is therefore valid only for $|u_c|$ well below $u_{\mathrm{sat}}$, as in the SOAEs of this study ($|u_c| \lesssim 7$~nm, Sec.~\ref{sec:analysis}).

In the smooth cochlea, $\gamma_0 = 1$ everywhere. To create a single place of reflection, $\gamma_0$ is given a step. The gain $\gamma_0$ is 1 basal to the step and 0.6 apical to the step. An abrupt change of the gain, even a small one, destabilizes a model~\cite{Elliott2007,Ku2008}. For the SOAEs, the step is at 10.5~mm (Secs.~\ref{sec:res_soae} and \ref{sec:analysis}). For the SFOAE, a smaller step, from 1 to 0.98, is placed near the peak of the probe response (Sec.~\ref{sec:res_sfoae}). The parameters are those of model 2 of Ref.~\cite{Sisto2019}.

\section{Numerical method}
\label{sec:method}

\subsection{Time integration}
\label{sec:overview}

\begin{figure}[t]
\includegraphics[width=0.75\linewidth]{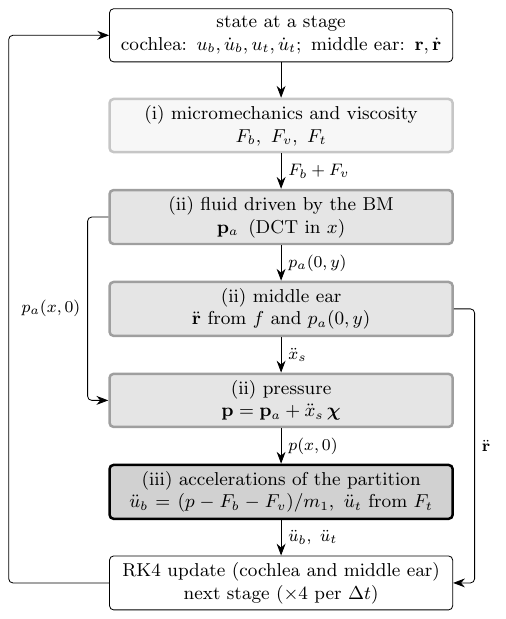}
\caption{\label{fig:dns}One stage of the time integration. Step (i)
computes the viscous damping $F_v$ and the micromechanical forces $F_b$ and $F_t$. Step (ii)
computes the pressure $\mathbf p_a$ driven by the BM, the acceleration $\ddot x_s$ of the stapes, and
the total pressure $\mathbf p$. Step (iii) computes the accelerations $\ddot u_b$ and $\ddot u_t$ of the
partition.}
\end{figure}

As in earlier time-domain solutions~\cite{Diependaal1987,Murakami2021}, the model is discretized in space and integrated in time. The state consists of velocity and displacement of the micromechanics. Because the fluid is incompressible, the pressure is not a state variable. The pressure follows from the state by a single linear solve. The time integration uses the classical fourth-order Runge--Kutta (RK4) method with $\Delta t = 5$~\textmu s, with the stimulus $f(t)$, the force on the probe diaphragm (Sec.~\ref{sec:incorporating}), sampled at intervals of $\Delta t/2$.

The pressure is coupled to the middle ear. The acceleration of the stapes
$\ddot{x}_s$ enters the condition at $x = 0$, Eq.~(\ref{eq:bc_stapes}), but
is not known when the pressure is computed. The acceleration $\ddot{x}_s$ depends on the fluid load on
the stapes, which in turn depends on the pressure. In the earlier
solution~\cite{Murakami2021}, the acceleration of the stapes was prescribed,
and this loop did not arise. Including the loop directly would put the
pressure into the condition at $x = 0$, whereas the fast solver of
Sec.~\ref{sec:fluid_sim} requires a condition there that is free of the
pressure~\cite{Murakami2021}. The loop is therefore closed by superposition.
The discretized problem is linear, so the pressure is split as
\begin{equation}
\mathbf{p} = \mathbf{p}_a + \ddot{x}_s\, \boldsymbol{\chi},
\label{eq:superposition_p}
\end{equation}
where $\mathbf{p}_a$ is the solution for $\ddot{x}_s = 0$, driven by $F_b$
and $F_v$ alone. The vector $\boldsymbol{\chi}$ is the solution for a unit
acceleration of the stapes with $F_b = F_v = 0$. The vector $\boldsymbol{\chi}$ does not depend on the
state and is computed once in advance.

Each stage consists of three steps, shown in Fig.~\ref{fig:dns}.
{\renewcommand{\theenumi}{\roman{enumi}}%
\renewcommand{\labelenumi}{(\theenumi)}%
\begin{enumerate}
\item $F_b$ and $F_t$ are evaluated from the state by Eqs.~(\ref{eq:Fb}) and (\ref{eq:Ft}), and $F_v$ by Eq.~(\ref{eq:Fv}).
\item The pressure is computed in three parts. First, $\mathbf{p}_a$ is
solved from $F_b$ and $F_v$ (Sec.~\ref{sec:fluid_sim}). Second, the middle
ear gives $\ddot{x}_s$ from the stimulus $f(t)$ and from $\mathbf{p}_a$ on
the face $x = 0$ (Sec.~\ref{sec:incorporating}). Third,
Eq.~(\ref{eq:superposition_p}) gives $\mathbf{p}$.
\item The accelerations of the partition follow from
Eqs.~(\ref{eq:eom_b}) and (\ref{eq:eom_t}). Those of the middle ear follow from
Sec.~\ref{sec:incorporating}.
\end{enumerate}}

\subsection{Fluid simulation}
\label{sec:fluid_sim}

Step (ii) of Fig.~\ref{fig:dns} needs the pressure before the BM acceleration is known. Substituting the motion of the BM, Eq.~(\ref{eq:eom_b}), into the condition at the BM, Eq.~(\ref{eq:bc_bm}), eliminates the BM acceleration. The result is the Robin condition
\begin{equation}
\frac{\partial p}{\partial y} - \frac{2\rho}{m_1}\, p = -\frac{2\rho}{m_1}\,(F_b + F_v) \quad \text{at } y = 0.
\label{eq:robin}
\end{equation}
Step (ii) obtains $\mathbf{p}_a$ by solving the Laplace equation, Eq.~(\ref{eq:laplace}), with this Robin condition, the conditions of Eqs.~(\ref{eq:bc_wall}) and (\ref{eq:bc_helico}), and Eq.~(\ref{eq:bc_stapes}) with $\ddot{x}_s = 0$, four times per time step. The vector $\boldsymbol{\chi}$ of Eq.~(\ref{eq:superposition_p}) is the solution with $F_b = F_v = 0$ and $\ddot{x}_s = 1$. The viscous damping $F_v$ is evaluated at each stage by a pair of discrete cosine transforms. The first gives the mode amplitudes $\hat{v}_q$ of the BM velocity in Eq.~(\ref{eq:modes}), and the second the sum of Eq.~(\ref{eq:Fv}).

The grid is uniform in $x$ and nonuniform in $y$. The steps in $y$ are fine at the BM, where the short waves near the characteristic place need them (Sec.~\ref{sec:viscous}). They grow geometrically toward the wall. The growth is kept mild because the waves reflected at the step and the motion of the oval window give the pressure a part that hardly decays across the scala. A grid that coarsens quickly, as designed for an exponentially decaying pressure~\cite{Murakami2019}, loses accuracy near the wall. The second derivative in $y$ is approximated by the three-point difference for nonuniform steps. The second derivative in $x$ is approximated by the central difference. The derivative conditions are imposed with ghost nodes, and the node at $x = L$ is excluded.

The difference operator in $y$ is independent of $x$. The $x$ direction is therefore diagonalized by the discrete cosine transform~\cite{Murakami2021}. The modes of this transform are the cosine modes $\cos k_q x$ of Eq.~(\ref{eq:modes}). The discrete problem thus splits into a tridiagonal system in $y$ for each cosine mode. Each stage needs only two outputs of the fluid (step (ii) of Fig.~\ref{fig:dns}), the pressure on the BM, $p_a(x,0)$, and the pressure on the face $x = 0$, $p_a(0,y)$. The first drives the BM. The second loads the stapes (Sec.~\ref{sec:incorporating}). The source of each cosine mode lies at $y = 0$, in the Robin condition. The pressure of a mode is therefore its source times a fixed profile in $y$. This profile is computed once in advance for each mode by a direct solution of its tridiagonal system. Each stage then needs only a pair of discrete cosine transforms and a product for each mode. The vector $\boldsymbol{\chi}$ is computed once in advance by the same solver.

The cosine transform in $x$ requires conditions at the ends in $x$ that do not contain the pressure~\cite{Murakami2021}. The Robin condition at the BM is solved within the tridiagonal system in $y$ of each cosine mode. If the middle ear were coupled directly, the condition at $x = 0$ would contain the pressure on the face through the load on the stapes. The condition at $x = 0$ would become a condition of the Robin type, like Eq.~(\ref{eq:robin}), and would couple the cosine modes. The superposition of Eq.~(\ref{eq:superposition_p}) keeps the condition at $x = 0$ free of the pressure.

The grid has $N_x = 2000$ points in $x$ and $N_y = 56$ in $y$, with a step of 10~\textmu m at the BM. The nodes in $x$ are $x_i = i\,\Delta x$, $i = 0, \ldots, N_x - 1$, with $\Delta x = L/N_x = 17.5$~\textmu m. The condition at $x = 0$ is imposed by a ghost node, and $p = 0$ at the excluded node $x = L$. The transform is then the discrete cosine transform of type III, and its inverse that of type II. The $N_y$ nodes in $y$ include both walls. The first 20 steps are 10~\textmu m, up to $y = 0.2$~mm. The following 35 steps grow by the factor 1.15, are capped at 0.5~mm, and are scaled together so that the last node lies at $y = H$. The steps thus range from 10~\textmu m to 0.47~mm. A 2D model needs a sufficiently fine grid~\cite{Murakami2018}. In preliminary simulations, the time-domain solution was therefore computed on successively refined grids. The estimated errors stayed within the range expected from the order of the difference scheme.

\subsection{Incorporating the micromechanics and the middle ear}
\label{sec:incorporating}

The micromechanics and the viscous damping enter the fluid simulation only
through their forces on the BM, $F_b$ and $F_v$, in
the Robin condition, Eq.~(\ref{eq:robin}). These forces follow from the state. The middle
ear enters through the acceleration of the stapes $\ddot{x}_s$ in
the condition at $x = 0$, Eq.~(\ref{eq:bc_stapes}). Step (ii) obtains
$\ddot{x}_s$ from the middle ear between $\mathbf{p}_a$ and $\mathbf{p}$
(Sec.~\ref{sec:overview}). This section describes that part.

Let $A_s$ be the area of the stapes. The fluid load on the stapes is
\begin{equation}
A_s \bar{p}_0 = A_s \left( \bar{p}_a + \bar{\chi}\, \ddot{x}_s \right),
\label{eq:load}
\end{equation}
where the bar denotes the mean over the face $x = 0$, taken with trapezoidal weights in $y$. The load uses the pressure difference $p$ itself, without a further factor. The second part of the load is proportional to
the acceleration of the stapes. This part acts as an added mass $A_s \bar{\chi}$
of the fluid and is moved to the mass of the stapes. Only the known part
$A_s \bar{p}_a$ then remains as a force on the middle ear.

The middle ear is the lumped model~\cite{LiuNeely2010} with the
parameters of that study. The model of the middle ear includes the probe diaphragm coupled through the air in the
ear canal. The three masses of the middle ear are the probe diaphragm, the umbo, and the
stapes. Their displacements $\mathbf{r} = [x_d, x_m, x_s]^T$ are positive
toward the cochlea. Their equations of motion are
\begin{equation}
\mathsf{M} \ddot{\mathbf{r}} + \mathsf{C} \dot{\mathbf{r}} + \mathsf{K} \mathbf{r}
= f(t)\, \mathbf{e}_d - A_s \bar{p}_a\, \mathbf{e}_s,
\label{eq:middle_ear}
\end{equation}
where $\mathsf{M}$, $\mathsf{C}$, and $\mathsf{K}$ are the mass, damping, and
stiffness matrices of the model~\cite{LiuNeely2010}, and $\mathbf{e}_d = [1, 0, 0]^T$ and
$\mathbf{e}_s = [0, 0, 1]^T$ select the probe diaphragm and the stapes.
The coupling to the present cochlea changes the model in
two ways. The added mass $A_s \bar{\chi}$ is added to the mass of the
stapes in $\mathsf{M}$. The force $-A_s \bar{p}_a$ acts on the stapes as the load of the
cochlea. The ear-canal pressure, which gives the OAE, is
\begin{equation}
P_d = K_c \left( A_d x_d - A_m x_m \right),
\label{eq:Pd}
\end{equation}
where $A_d$ and $A_m$ are the areas of the probe diaphragm and the eardrum,
and $K_c$ is the pressure of the air in the ear canal per unit volume displacement. The values are $A_s = 6.25$~mm$^2$, $A_d = 75$~mm$^2$, $A_m = 50$~mm$^2$, and $K_c = 8.5 \times 10^{10}$~Pa/m$^3$~\cite{LiuNeely2010}.
The middle ear is integrated in the same RK4 stages as the cochlea.

\section{Emissions from the model}
\label{sec:results}

\subsection{SFOAE suppression tuning}
\label{sec:res_sfoae}

\begin{figure}[t]
\includegraphics[width=\linewidth]{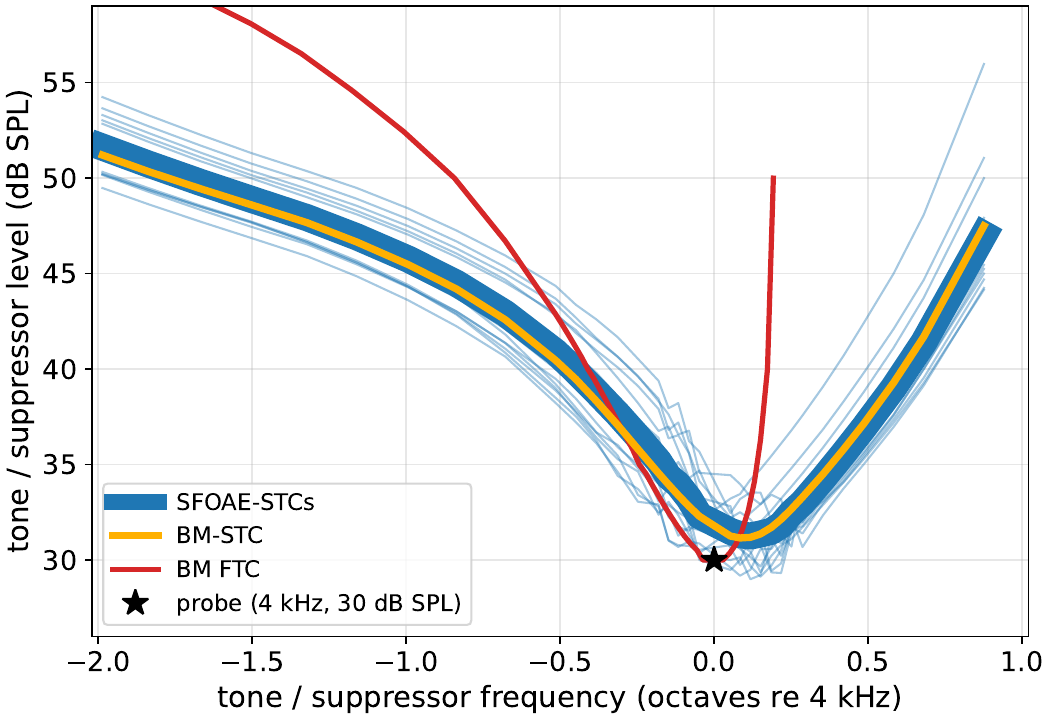}
\caption{\label{fig:stc}SFOAE suppression tuning curve (SFOAE-STC), BM
suppression tuning curve (BM-STC), and BM frequency tuning curve (FTC).
The star marks the probe.
The thin lines are the SFOAE-STCs for the positions of the step of
$\gamma_0$, and the thick line is their mean. The BM curves are those of the smooth cochlea at the
peak of the probe response.}
\end{figure}

\begin{table}[t]
\caption{\label{tab:tuning}Tip, sharpness, and slopes of the tuning curves of Fig.~\ref{fig:stc}. The tip frequency is in octaves re the probe frequency $f_p$. $Q_{10}$ is $f_p$ divided by the bandwidth 10~dB above the tip.}
\begin{ruledtabular}
\begin{tabular}{lrrrrr}
Curve & \multicolumn{2}{c}{Tip} & $Q_{10}$ & \multicolumn{2}{c}{Slope (dB/oct)} \\
 & (oct) & (dB SPL) & & Below & Above \\
\hline
SFOAE-STC & $+0.118$ & $31.3$ & $1.10$ & $-13$ & $13$ \\
BM-STC & $+0.084$ & $31.2$ & $1.11$ & $-13$ & $13$ \\
BM FTC & $-0.037$ & $30.0$ & $2.67$ & $-26$ & $63$ \\
\end{tabular}
\end{ruledtabular}
\end{table}

Before the spontaneous emissions, Fig.~\ref{fig:stc} compares the tuning of the SFOAE with that of the BM. The SFOAE-STC is obtained by the suppression method of Ref.~\cite{Charaziak2014}, with the probe and suppressor frequencies of that study. The SFOAE-STC is the suppressor level at which the residual of $P_d$ at the probe frequency $f_p$ falls to a criterion. The step of $\gamma_0$ provides the reflection site that the SFOAE requires~\cite{Shera1999,ZweigShera1995}. The step is placed at the peak of the probe response, 12.74~mm for the probe of 4~kHz at 30~dB SPL. The STCs depend on the exact position of the step, so those for several positions are averaged. The step, abrupt at one grid point, is placed at 22 positions from 12.18 to 13.02~mm, at every grid point from 12.18 to 12.39~mm and at every fourth from 12.46 to 13.02~mm. The criterion of the SFOAE-STC is a residual of $-36$~dB SPL. The BM-STC and the BM FTC are computed in the smooth cochlea at the peak of the probe response. The criterion of the BM-STC is a residual of the BM velocity 15~dB below its maximum over the suppressor frequencies, the depth of the BM suppression at the thresholds of the SFOAE-STC. The criterion of the BM FTC is the BM velocity of the probe alone. Each tone is calibrated in $P_d$ and turned on with a 5-ms raised-cosine ramp. After 0.1~s, two rectangular windows of 0.1~s are analyzed. The second gives the response, and the first checks that the response is steady. The thresholds are interpolated linearly in level. The slopes are fitted over 0.3~octave below and above the tip, and $Q_{10}$ is computed from the mean SFOAE-STC.

Table~\ref{tab:tuning} summarizes the shapes of the three curves. The BM FTC has a sharp tip near $f_p$. The slope of the BM FTC above the tip is steeper than that below. This asymmetry is that of the measured BM tuning curves~\cite{RoblesRuggero2001}.

The SFOAE-STC and the BM-STC have the same shape. Their tips, sharpnesses, and slopes agree. Both are less than half as sharp as the BM FTC, in particular above $f_p$. The SFOAE suppression thus reflects the suppression of the BM response at the place of the probe rather than the frequency tuning of the BM. The same result was found in measurements, where the SFOAE-STCs were broader than the compound-action-potential STCs~\cite{Charaziak2014}. The SFOAE-STC of the model is broader than the measured one, but the trend is the same. A 2D model without a middle ear gave the same result~\cite{MurakamiFuji2022}.

These results support the model for the emissions that arise by reflection. The BM tuning of the model has the shape found in the measurements. The SFOAE generated by the reflection at the step follows the BM at the place of the reflection, as in the measurements of the SFOAE. The model thus relates the emission in the ear canal to the motion of the BM at the reflection site. The spontaneous emissions below arise from the same kind of reflection.

\subsection{Spontaneous emissions}
\label{sec:res_soae}

\begin{figure*}[ht]
\includegraphics[width=\textwidth]{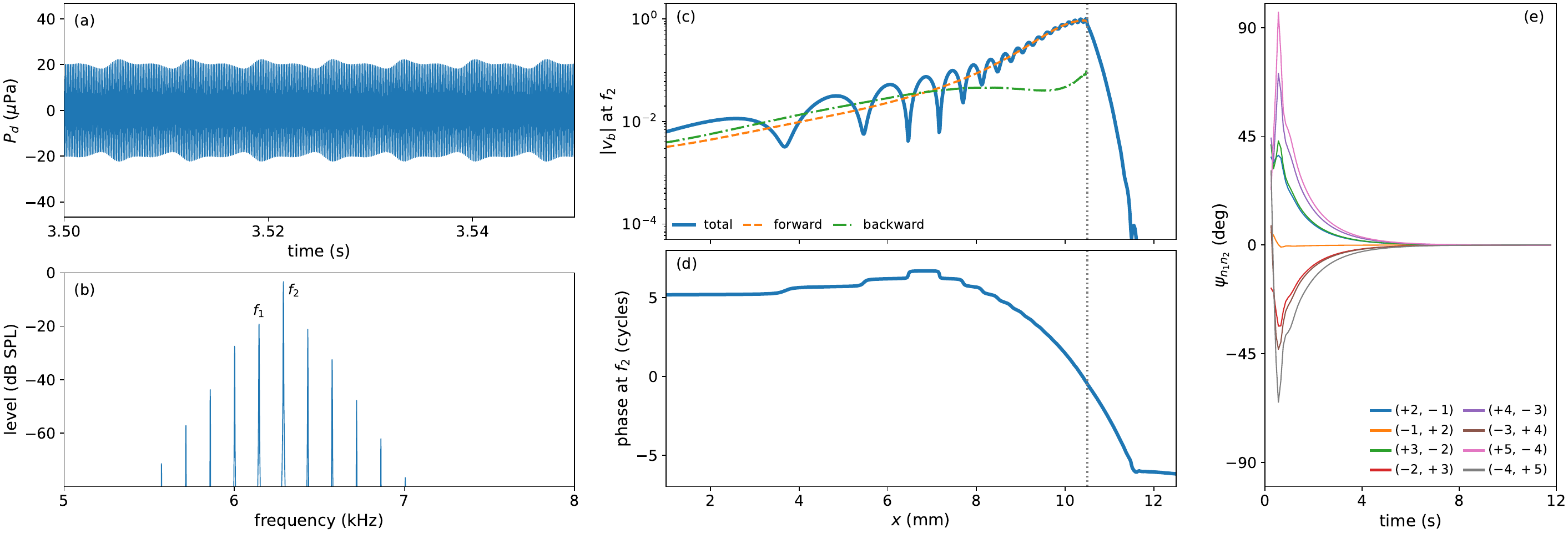}
\caption{\label{fig:soae}Spontaneous emission for the single step of $\gamma_0$.
The initial BM velocity is uniform, 1~\textmu m/s, and the rest of the state is zero. No stimulus is applied.
(a) Waveform of the ear-canal pressure $P_d$ band-passed to 2--20~kHz.
(b) Spectrum of $P_d$ over the last 1~s of a 4-s solution, with a Hann window. The two strongest lines are labeled $f_2$ and $f_1$.
(c) Amplitude of the BM velocity at the line $f_2$ in the steady state of the time-domain solution, 11.5--12~s, normalized by its maximum over $x$. The dashed and dash-dotted lines are the forward and backward waves of this velocity, shown basal to the step.
(d) Phase of the BM velocity at the line $f_2$, zero at the step.
(e) Phase differences $\psi_{n_1 n_2}$ of Eq.~(\ref{eq:psi}) of the intermodulation products relative to their final values, labeled by $(n_1, n_2)$.
The dotted lines in (c) and (d) mark the step.}
\end{figure*}

Figures~\ref{fig:soae}(a) and \ref{fig:soae}(b) show the ear-canal pressure $P_d$ without stimulus. The emission is a group of equally spaced lines. The main line is denoted by $f_2$. The next strongest line is denoted by $f_1$.

The emissions of the model share three properties with measured SOAEs. First, each line is narrow-band. Human SOAEs are narrow-band signals whose amplitude and period fluctuate by only a few percent or less~\cite{vanDijkWit1990}. The lines of the model, computed without noise, do not fluctuate at all. Second, the emission arises at the edge of a region of reduced gain. In a guinea pig, an SOAE was accompanied by a spontaneous oscillation of the BM, in an ear whose sensitivity was reduced at frequencies below that of the emission~\cite{Nuttall2004}. The model likewise produces the BM oscillation together with the ear-canal pressure. Third, the spectrum contains multiple lines whose phases lock. Human SOAE spectra may contain multiple peaks, and in equally spaced triplets the phases lock as those of a cubic distortion product~\cite{vanDijkWit1998}.

The BM oscillation of the main line $f_2$ is a standing wave, as shown in Figs.~\ref{fig:soae}(c) and \ref{fig:soae}(d). These panels show the steady state of the solution. Near the step, the pattern is a traveling wave. On the basal side, the phase is constant. To separate the two waves, the BM velocity $v_b(x, t) = \dot{u}_b(x, t)$ is transformed in time and space to $v_b(k, f)$, where $k$ is the wavenumber. The transforms use the kernels $e^{-2\pi i f t}$ and $e^{-ikx}$, the time window is a Hann window, and the space transform covers the whole length without a window. At the frequency of the main line, the components with $k < 0$ form the forward wave and those with $k > 0$ the backward wave. Each is transformed back to $x$. Basal to the step, the standing wave consists of these forward and backward waves, as shown in Fig.~\ref{fig:soae}(c). SOAEs have been interpreted as such standing waves~\cite{Shera2003}.

As shown in Fig.~\ref{fig:soae}(b), the lines are spaced by $\Delta = f_2 - f_1$.
Their frequencies are thus $n_1 f_1 + n_2 f_2$, with integers $n_1$ and $n_2$.
If a line is an intermodulation product of the lines $f_1$ and $f_2$, its
phase follows from their phases $\phi_1$ and $\phi_2$ as
$n_1\phi_1 + n_2\phi_2$. The actual phase $\phi_{n_1 n_2}$ of each line is
compared with this phase by the difference
\begin{equation}
\psi_{n_1 n_2} = \phi_{n_1 n_2} - (n_1\phi_1 + n_2\phi_2).
\label{eq:psi}
\end{equation}
This difference stays constant for an intermodulation product and drifts for an independent oscillation. The phases are those of the Hann-windowed Fourier transform of $P_d$ at $n_1 f_1 + n_2 f_2$, in windows of 0.5~s shifted by 0.1~s. The frequencies $f_1$ and $f_2$ are the peaks of the spectrum over 6--12~s. In the solution, $\psi_{n_1 n_2}$ of every line settles to a constant, as shown in Fig.~\ref{fig:soae}(e). The lines are thus intermodulation products of the main line $f_2$ and the line $f_1$.

The levels of the emissions are not fitted. Without stimulus, the only nonlinearity is the saturation of Eq.~(\ref{eq:gamma}). The solution therefore scales with $u_{\mathrm{sat}}$, taken from model 2 of Ref.~\cite{Sisto2019}. The frequencies, the spacing, and the stability of the lines do not depend on $u_{\mathrm{sat}}$.

\section{Analysis of the emissions}
\label{sec:analysis}

\subsection{The main line and the saturation}
\label{sec:mainline}

The analysis starts from the main line $f_2$ of the emission, the strongest line, as shown in Fig.~\ref{fig:soae}(b). The other lines are the intermodulation products with the line $f_1$, as shown in Fig.~\ref{fig:soae}(e). The oscillation of the main line alone is therefore computed first by harmonic balance, as described in Appendix~\ref{app:pump}. The saturation of Eq.~(\ref{eq:gamma}) limits the amplitude of the main line.

The harmonic balance calculates $u_c = A\cos\omega t$ at each place, where $A$ is the amplitude of the main line and $\omega = 2\pi f_2$ its angular frequency. The other lines are much weaker, so the main line alone sets the gain. The factor of the gain in Eq.~(\ref{eq:gamma}) is then
\begin{equation}
1 - \frac{u_c^2}{u_{\mathrm{sat}}^2}
= \underbrace{1}_{\text{linear}}
- \underbrace{\frac{A^2}{2u_{\mathrm{sat}}^2}}_{\text{cross-saturation}}
- \underbrace{\frac{A^2}{2u_{\mathrm{sat}}^2}\cos 2\omega t}_{\text{pump}}.
\label{eq:uc2}
\end{equation}
Equation~(\ref{eq:uc2}) shows the factor of the gain alone. The active force is this factor times $c_4\dot{u}_c + k_4 u_c$. The exact linearization in Appendix~\ref{app:pump} differentiates this product, so the damping and the stiffness are lowered by different amounts.

The three terms correspond to the three kinds of eigenvalues in Secs.~\ref{sec:linear} and \ref{sec:quasilinear}. The first term gives the linear model, analyzed in Sec.~\ref{sec:linear}. The second term is always negative. The second term thus lowers the mean gain where the main line is large. Every other line sees this lower gain. This lowering of the mean gain is the \textit{cross-saturation}. 

\subsection{The pump}
\label{sec:pump}

\begin{figure}[t]
\includegraphics[width=\linewidth]{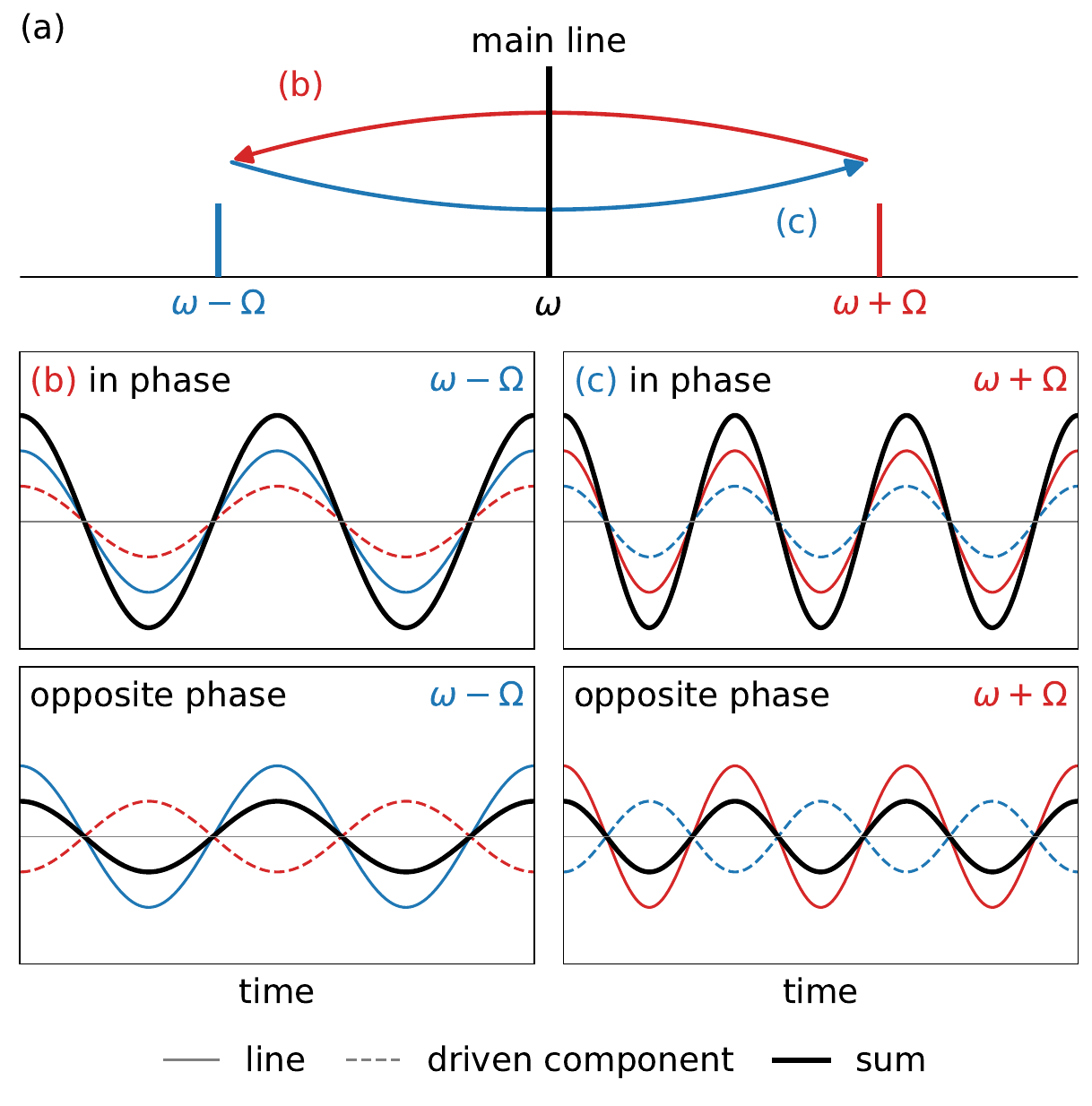}
\caption{\label{fig:pumpsketch}(a) The pump at $2\omega$ and the two lines at
$\omega \pm \Omega$. Each line drives its mirror image about the main line.
(b) The line at $\omega - \Omega$ (solid), the component driven by the line
at $\omega + \Omega$ through the pump (dashed), and the sum of the two
(thick). The component is in phase with the line (top) and in opposite
phase (bottom). (c) The same at $\omega + \Omega$, with the component driven
by the line at $\omega - \Omega$.}
\end{figure}

The third term of Eq.~(\ref{eq:uc2}) oscillates at $2\omega$. The third term thus modulates the gain in time. The factor $\cos 2\omega t$ is this oscillation of the gain. Consider a line shifted by $\Omega$ from the main line, $\cos[(\omega + \Omega)t]$. The active force on this line is the gain times the line. The oscillation of the gain multiplies the line as
\begin{equation*}
\cos 2\omega t \, \cos[(\omega + \Omega)t]
= \frac{1}{2}\cos[(\omega - \Omega)t]
	+ \frac{1}{2}\cos[(3\omega + \Omega)t].
\end{equation*}
The component at $3\omega + \Omega$ is far from the lines and is neglected. The other component is at the difference frequency $2\omega - (\omega + \Omega) = \omega - \Omega$. The line at $\omega + \Omega$ thus drives a line at $\omega - \Omega$. In the same way, a line at $\omega - \Omega$ drives one at $\omega + \Omega$. The oscillation of the gain thus couples two lines placed symmetrically about the main line, as shown in Fig.~\ref{fig:pumpsketch}(a). This oscillating gain is the pump introduced in Sec.~\ref{sec:intro}. The coupling is the four-wave mixing of a saturable gain~\cite{Agrawal1988}.

Whether the drive by the pump raises or lowers a line depends on the phase. The line at $\omega \pm \Omega$ drives a component at $\omega \mp \Omega$. If each component is in phase with the line at the same frequency, the sum is larger than the line alone, as shown in the top of Figs.~\ref{fig:pumpsketch}(b) and \ref{fig:pumpsketch}(c). The component then amplifies the line. If the component is in opposite phase, the sum is smaller, as shown in the bottom of Figs.~\ref{fig:pumpsketch}(b) and \ref{fig:pumpsketch}(c). The component then suppresses the line. An amplified line drives the other line more strongly. Through this loop, shown in Fig.~\ref{fig:pumpsketch}(a), the two lines amplify each other in phase and suppress each other in opposite phase. If this mutual amplification exceeds the decay of each line alone, the pair grows.

The pump acts on a pair at any $\Omega$. The decay of each line alone, however, depends on the frequency of the line. A line decays slowly only near a weakly damped mode of the cochlea. A pair can therefore grow only at a particular $\Omega$, where one of the two lines lies near such a mode. In the model, the weakly damped mode lies just above the line $f_1$.

\subsection{Linear eigenvalues}
\label{sec:linear}

\begin{figure}[t]
\includegraphics[width=\linewidth]{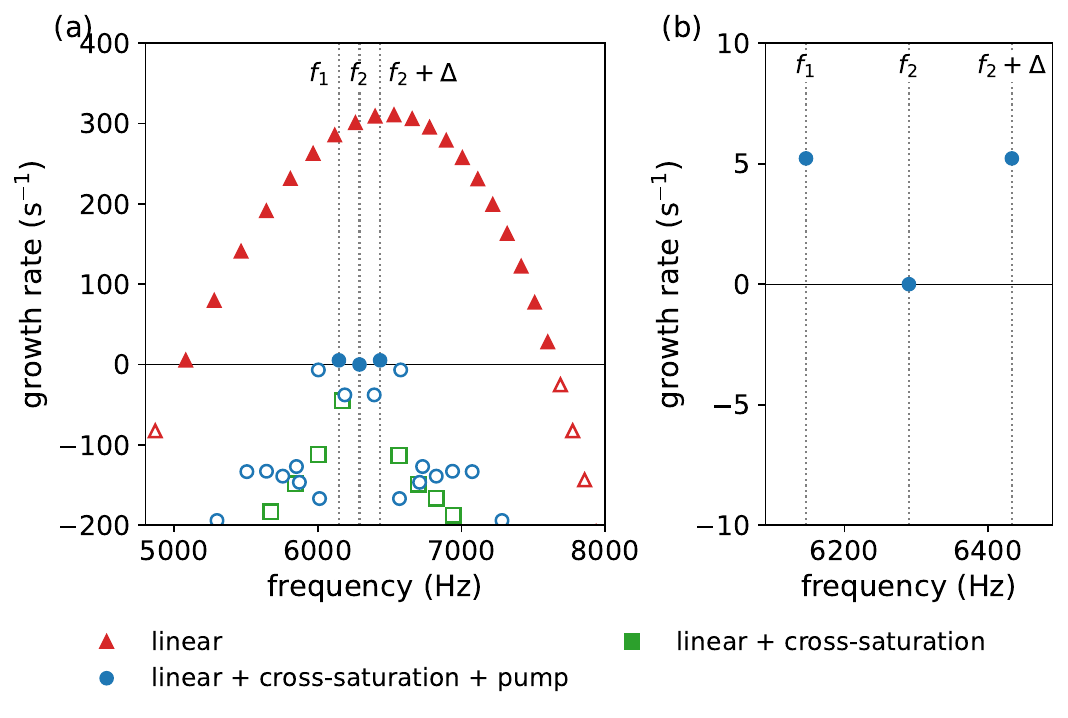}
\caption{\label{fig:pump}Linear eigenvalues of $\mathsf{J}$ (triangles) and quasilinear eigenvalues of
Eq.~(\ref{eq:pair}) with the cross-saturation added (squares) and with the pump also added (circles).
Filled markers are unstable, and open markers stable. Panel (b) magnifies the pair and the main line of (a). Each
eigenvalue with the pump consists of two lines placed symmetrically about
the main line and is plotted at both. The pair closest to zero is unstable. The circle at the main line with
$s = 0$ is the main line itself. The dotted lines
mark the main line and the two lines of the pair.}
\end{figure}

Without stimulus ($f = 0$), the state of rest is an equilibrium of the model. The stability of this state decides whether an SOAE arises, and at which frequencies. As in state-space models~\cite{Elliott2007,Ku2008,Ku2009}, the stability is analyzed by the eigenvalues of the model linearized about rest. Here the model includes a 2D fluid and the middle ear. Since the bracket of Eq.~(\ref{eq:gamma}) differs from 1 only at second order in $u_c$, the linearized model is the model with $\gamma = \gamma_0(x)$.

The discretized model is a system of first-order equations $\dot{\mathbf{w}} = \mathbf{F}(\mathbf{w}, f)$ for a state vector $\mathbf{w}$ that contains the displacements and velocities of the BM and the TM at the $N_x$ grid points. The vector also contains those of the probe diaphragm, the umbo, and the stapes. The model thus has $4N_x + 6$ state variables. Without stimulus and linearized about rest, the system is
\begin{equation}
\dot{\mathbf{w}} = \mathsf{J}\, \mathbf{w}.
\label{eq:linear_system}
\end{equation}
The matrix $\mathsf{J}$ is the Jacobian matrix of $\mathbf{F}$ at rest and is the state matrix of the linearized model~\cite{Elliott2007,Ku2008,Ku2009}. Appendix~\ref{app:J} describes the computation of the Jacobian $\mathsf{J}$.

The eigenvalues $s_m = \sigma_m + i\omega_m$ of $\mathsf{J}$ give the modes of the model. For $\omega_m > 0$, $\omega_m/2\pi$ is the frequency of mode $m$ and $\sigma_m$ its growth rate. If all $\sigma_m < 0$, the state of rest is stable and the model has no SOAE. If any $\sigma_m > 0$, the state of rest is unstable. The cubic term of Eq.~(\ref{eq:gamma}) then limits the oscillation. The amplitude of the oscillation is found in the time domain.

The linear eigenvalues of $\mathsf{J}$ show several unstable modes, as the triangles in Fig.~\ref{fig:pump}(a) show. The unstable mode nearest to the main line $f_2$ is called mode 2, and that nearest to the line $f_1$ is called mode 1.

\subsection{Quasilinear eigenvalues}
\label{sec:quasilinear}

The linear eigenvalues of $\mathsf{J}$ do not include the cross-saturation and the pump by the main line. The model is linearized about the oscillation of the main line. The pair is small compared with the main line, so the linearization holds. The linearized model follows from Eq.~(\ref{eq:linear_system}) with the gain factor of Eq.~(\ref{eq:uc2}).

The first term of Eq.~(\ref{eq:uc2}) gives $\mathsf{J}$. The second term is constant in time. This constant lowering of the gain turns $\mathsf{J}$ into another fixed matrix $\mathsf{J}_0$, as described in Appendix~\ref{app:pump}. The third term, in contrast, varies in time as $\cos 2\omega t$. This term cannot be put into a fixed matrix in the same way. The third term is therefore kept as a separate matrix multiplied by the time factor. By Euler's formula, $\cos 2\omega t = (e^{2i\omega t} + e^{-2i\omega t})/2$. The third term thus gives a matrix $\mathsf{P}$ multiplied by $e^{2i\omega t}$ and the conjugate matrix $\bar{\mathsf{P}}$ multiplied by $e^{-2i\omega t}$. The gain is real, so the two matrices $\mathsf{P}$ and $\bar{\mathsf{P}}$ are conjugate to each other. Appendix~\ref{app:pump} gives $\mathsf{J}_0$ and $\mathsf{P}$ from the linearization of the cubic term, in Eqs.~(\ref{eq:J0}) and (\ref{eq:P}).

The linearized model is thus
\begin{equation}
\dot{\mathbf{w}} = \mathsf{J}_0\,\mathbf{w} + \left(\mathsf{P}\,e^{2i\omega t} + \bar{\mathsf{P}}\,e^{-2i\omega t}\right)\mathbf{w}.
\label{eq:lin_main}
\end{equation}
The second term is the pump. The pump couples only the two lines at $\omega \pm \Omega$, as described in Sec.~\ref{sec:pump}. Solutions are sought as two lines that grow together at the rate $\sigma$. The line at $\omega + \Omega$ is then $e^{\sigma t} e^{i(\omega + \Omega)t} = e^{(s + i\omega)t} \equiv e_+$, with $s = \sigma + i\Omega$. The conjugate of the line at $\omega - \Omega$ is $e^{\sigma t} e^{-i(\omega - \Omega)t} = e^{(s - i\omega)t} \equiv e_-$, with the same $s$. Equation~(\ref{eq:lin_main}) is linear and has real coefficients. The real part of a complex solution is thus also a solution. The state is therefore sought as the complex solution
\begin{equation}
\mathbf{w} = \mathbf{c}_+ e_+ + \mathbf{c}_- e_-.
\label{eq:trial}
\end{equation}
Here $\mathbf{c}_+$ and $\mathbf{c}_-$ are constant complex vectors over the state. Equation~(\ref{eq:trial}) is substituted into Eq.~(\ref{eq:lin_main}). The time derivative and the pump act on the two terms of Eq.~(\ref{eq:trial}) as
\begin{align*}
\frac{d}{dt}\left(\mathbf{c}_\pm e_\pm\right) &= (s \pm i\omega)\,\mathbf{c}_\pm e_\pm,\\
	e^{\pm 2i\omega t}\, e_{\mp} &= e_{\pm}.
\end{align*}
The pump thus turns each line into the other. The other products of the pump oscillate near $3\omega$ and are neglected. The factors $e_+$ and $e_-$ oscillate at different frequencies. The terms with each factor must therefore agree on the two sides of Eq.~(\ref{eq:lin_main}),
\begin{align*}
(s + i\omega)\,\mathbf{c}_+ &= \mathsf{J}_0\,\mathbf{c}_+ + \mathsf{P}\,\mathbf{c}_-,\\
(s - i\omega)\,\mathbf{c}_- &= \mathsf{J}_0\,\mathbf{c}_- + \bar{\mathsf{P}}\,\mathbf{c}_+.
\end{align*}
Moving $\pm i\omega\,\mathbf{c}_\pm$ to the right sides gives
\begin{equation}
s \begin{pmatrix} \mathbf{c}_+ \\ \mathbf{c}_- \end{pmatrix} =
\begin{pmatrix}
\mathsf{J}_0 - i\omega & \mathsf{P} \\
\bar{\mathsf{P}} & \mathsf{J}_0 + i\omega
\end{pmatrix}
\begin{pmatrix} \mathbf{c}_+ \\ \mathbf{c}_- \end{pmatrix}.
\label{eq:pair}
\end{equation}
The oscillating factors have disappeared, and the coefficients are fixed matrices.
The eigenvalues $s$ of Eq.~(\ref{eq:pair}) are called the \textit{quasilinear eigenvalues}. Each eigenvector consists of two lines placed symmetrically about the main line. Without the pump ($\mathsf{P} = 0$), the two lines decouple. The quasilinear eigenvalues are then those of $\mathsf{J}_0$, measured from the main line. One eigenvalue with the pump is $s = 0$. This eigenvalue is the main line itself, shifted in time.

Without the pump, all the modes decay, as the squares in Fig.~\ref{fig:pump} show. Mode 1 also decays and lies above the line $f_1$. The mode of the main line decays too, because $\mathsf{J}_0$ applies the full cross-saturation to this mode. The main line itself sees only half of this lowering, the self-saturation (Appendix~\ref{app:pump}). Without the pump, the saturation thus stabilizes all the modes other than the main line. The saturation alone favors a single line.

The pump destabilizes one pair of lines, as shown in Figs.~\ref{fig:pump}(a) and \ref{fig:pump}(b). The pair lies at the line $f_1$ and at the mirror image of $f_1$ about the main line, the line $f_2 + \Delta$. All the other pairs decay. Without the pump, each line of the pair decays. The growth of the pair is thus the effect of the pump. The pump also equalizes the spacing of the lines.

\begin{figure}[t]
\includegraphics[width=\linewidth]{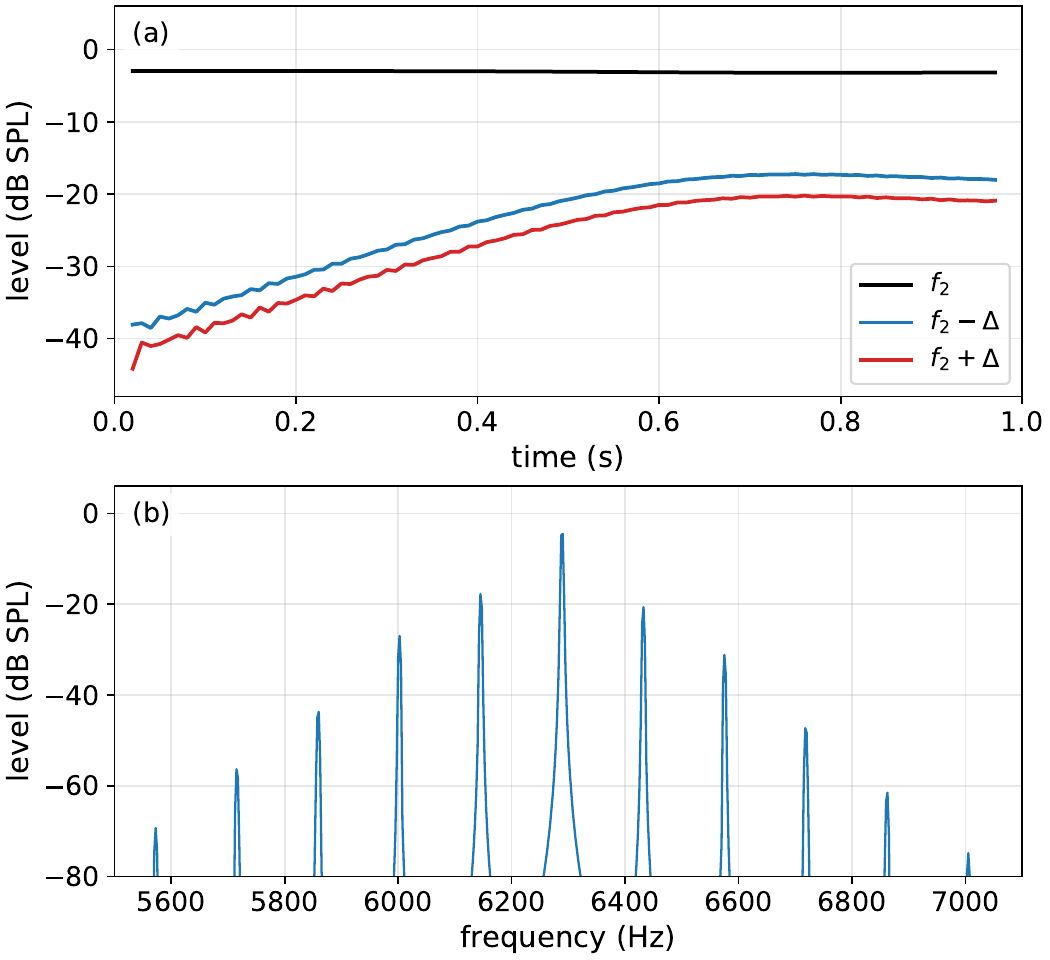}
\caption{\label{fig:growth}Time-domain solution started from the mode of the
main line alone. (a) Levels of the main line $f_2$ and of the sidebands
$f_2 \pm \Delta$ in the ear-canal pressure $P_d$, from a 40-ms Hann window
shifted by 10~ms. (b) Spectrum of $P_d$ over
0.6--1.0~s.}
\end{figure}

The time-domain solution confirms this pair. The model is solved from an initial condition that contains only the mode of the main line. This mode is the eigenvector nearest to $f_2$ of the model with the self-saturation of Eq.~(\ref{eq:df_single}), scaled to the peak BM velocity of the main line. The initial condition is not the harmonic-balance solution, so the main line first adjusts to its steady state. Two sidebands at $f_2 \pm \Delta$ then grow together out of the main line, as Fig.~\ref{fig:growth}(a) shows. The single main line turns into a group of lines, as shown in Fig.~\ref{fig:growth}(b). The lower sideband lies at the frequency of the line $f_1$. The growth rate of the sidebands is close to that of the pair. The oscillation of the main line alone is therefore unstable to modulation. The other lines of the emission are phase-locked intermodulation products of the main line and the line $f_1$ (Sec.~\ref{sec:res_soae}). The line $f_1$ itself thus arises from the main line alone.

\section{Discussion}
\label{sec:discussion}

\subsection{The pump creates the lines}
\label{sec:mc}

The pump destabilizes one pair of lines, and this pair produces the group of
lines. The unstable pair is shown by the filled circles in
Fig.~\ref{fig:pump}(b). The pair lies at the line $f_1$ and at the line
$f_2 + \Delta$. The other lines of the emission are phase-locked
intermodulation products of the main line and the pair, as
Fig.~\ref{fig:soae}(e) shows.

Without the pump, the cross-saturation stabilizes all the modes, as the
squares in Fig.~\ref{fig:pump}(a) show. Several modes are unstable in the
linear eigenvalues, shown by the triangles, but all of these modes decay
with the cross-saturation. That not all unstable modes become SOAEs was
found in state-space models~\cite{Ku2009}. The cross-saturation gives a
reason for this finding.

A pump generated by the main line itself has not been stated for a
cochlear model, but
similar mechanisms are known in other oscillators. In earlier measurements, an external
low-frequency tone modulated the amplitude of SOAEs and produced
sidebands~\cite{BianWatts2008}. In the model, the main line itself modulates
the gain of the active element. In delayed-feedback oscillators, modes that are
damped alone are likewise excited on the background of the main
oscillation~\cite{Emelianova2014}. In optical microresonators, four-wave
mixing equalizes the spacing of the sidebands against the dispersion, by a
nonlinear refractive index~\cite{Kippenberg2011}. In quantum cascade
lasers with a fast saturable gain, four-wave mixing locks the modes into a
comb~\cite{Hugi2012,Khurgin2014}.

\subsection{The two explanations of multiple SOAEs}
\label{sec:two}

In the model, the two explanations of multiple SOAEs hold together. The
standing-wave view accounts for the origin of the main line and for the
spacing~\cite{Shera2003,Shera2022}. The main line is a standing wave between
the step and the stapes, as the forward and backward waves in
Fig.~\ref{fig:soae}(c) show. The pair grows near the neighboring linear
modes, as Fig.~\ref{fig:pump}(b) shows, and the pump equalizes the spacing
of the lines in Fig.~\ref{fig:soae}(b). The intermodulation view accounts
for the nature of the other lines. The phases of the other lines lock as
those of intermodulation products, as Fig.~\ref{fig:soae}(e) shows. The
phases of equally spaced triplets of human SOAEs lock in the same
way~\cite{vanDijkWit1998}.

The lines grow from one oscillation, not from independent oscillations.
Locked phases alone do not show a single oscillation, since coupled
independent oscillators also lock~\cite{Gelfand2010,WitVanDijk2012}. Without
the pump, all the modes decay, as Fig.~\ref{fig:pump}(a) shows. The
time-domain solution started from the main line alone produces the
sidebands, as Fig.~\ref{fig:growth} shows. In the model, the
lines are therefore neither independent resonances nor intermodulation
products of independent oscillations~\cite{Ku2009}.

The mechanism concerns a group of lines from one standing wave, not SOAE
spectra in general. Most neighboring human SOAEs are not related by
intermodulation~\cite{Bergevin2025}. In barn owls, by contrast, consecutive
SOAEs are often nearly equally spaced. This near-equal spacing has been
attributed either to intermodulation or to a characteristic delay, such as
that of a standing wave~\cite{Bergevin2025}. The mechanism of the model
would allow both at once. The avian basilar papilla, however, differs from
the mammalian cochlea of the model, with a thick tectorial plate and many
hair cells across its width. Whether the papilla supports such a standing
wave is not addressed by the model.

\subsection{Standing wave or oscillator}

Spectra with several lines, raised in Sec.~\ref{sec:intro} as a test of the
two views of SOAEs, do not by themselves decide between the two views. In
the model, the whole standing wave acts as one self-sustained oscillation,
and the lines come from this one oscillation, as shown in
Sec.~\ref{sec:two}.
In this respect, the model is close to the oscillator view. The neighboring
resonances that set the spacing, however, come from the standing wave.
Whether a local oscillator without such resonances could produce the same
lines is not known. The model thus does not exclude other views of SOAEs,
such as coupled Hopf oscillators~\cite{Hudspeth2010,Fruth2014} or
fluid-coupled elements without a regular reflection site~\cite{Wit2026}.

\subsection{Assumptions and limitations of the model}

The distribution of the gain of the active element in the model differs
from that in real ears. The model has a single step in the gain, as at the
edge of a region of hearing loss~\cite{Nuttall2004}. In real ears, distributed roughness is
the prevailing explanation of the reflection~\cite{ZweigShera1995,Shera2003},
and a single site has been judged insufficient to account for spontaneous
BM oscillations~\cite{deBoerNuttall2006}. In the model, the distribution of the
gain can be set freely, and the step places the reflection at a known
position. A single step yields a group of lines from one place, not the
many SOAEs of a real ear~\cite{Burns2009,WitBell2024}. With distributed
roughness, the resonances would not be equally spaced. The pump still
couples only lines placed symmetrically about the main line, and a pair
grows when one of its lines lies near a weakly damped mode
(Sec.~\ref{sec:pump}). The mirror line need not lie on a resonance. Whether
the pair still grows when roughness shifts and damps the neighboring modes
irregularly was not examined.

The round-trip delay of the model is longer than in human ears. The delay
computed from the line spacing $\Delta$ is about twice the OAE group delay of
human ears at high frequencies~\cite{SheraGuinan2003}. A shorter delay would
double the spacing of the resonances and move the neighboring modes farther
from the main line. Whether the pump then still overcomes the
cross-saturation was not examined.

The saturation is simplified in two ways. The active element saturates as a cubic
and generates no even harmonics~\cite{Cheatham2018}. An asymmetric
saturation, such as that of the transduction of the outer hair cells, would
also modulate the gain at $\omega$ and shift its mean. The modulation at
$\omega$ couples a line at $\omega + \Omega$ to one at $\Omega$, far from the
main line, whereas the pump at $2\omega$ couples the pair about the main
line. How the asymmetry changes the balance between the cross-saturation and
the pump was not examined. The main line is
treated by a describing function, which holds because a single line
dominates~\cite{Ku2009}. The quasilinear eigenvalues describe the onset of
the pair. In the steady state, the line $f_1$ is no longer small compared
with the main line, and the pair then also saturates the main line.
The steady state is therefore given by the time-domain solution, not by the
eigenvalues.

\section{Conclusions}
\label{sec:conclusion}

A group of equally spaced SOAE lines can arise from a single cochlear
standing wave. The cross-saturation by the main line suppresses the other lines. The
pump destabilizes a pair of lines near the neighboring resonances. The other
lines are intermodulation products of the main line and the pair. In this
mechanism, the standing-wave view and the intermodulation view hold
together.

\begin{acknowledgments}
This work was supported by JSPS KAKENHI Grant Number JP26K14881.
A generative AI tool, Claude (Anthropic), was used to assist with language editing and computer
programming. The author reviewed and edited all output and takes full
responsibility for the content of this article.
\end{acknowledgments}

\section*{Author Declarations}
\subsection*{Conflict of Interest}
The author has no conflicts to disclose.

\section*{Data Availability}
The data that support the findings of this study are available from the
author upon reasonable request.

\appendix

\section{State matrix}
\label{app:J}

This appendix describes how the state matrix $\mathsf{J}$ of
Eq.~(\ref{eq:linear_system}) is computed. The linearization is made on the discretized model of
Sec.~\ref{sec:overview}, written as a system of first-order equations for
the state vector
\begin{equation}
\mathbf{w} = \left[ \mathbf{u}_b^T,\ \dot{\mathbf{u}}_b^T,\ \mathbf{u}_t^T,\
\dot{\mathbf{u}}_t^T,\ \mathbf{r}^T,\ \dot{\mathbf{r}}^T \right]^T
\label{eq:state}
\end{equation}
of $4N_x + 6$ components. The vectors $\mathbf{u}_b$ and $\mathbf{u}_t$
contain the displacements at the grid points of the partition. The vector
$\mathbf{r}$ contains the displacements of the probe diaphragm, the umbo,
and the stapes of Sec.~\ref{sec:incorporating}. In this state, the discretized
model is the system of ordinary differential equations integrated in time
(Sec.~\ref{sec:overview}),
\begin{equation}
\dot{\mathbf{w}} = \mathbf{F}(\mathbf{w}, f),
\label{eq:ode}
\end{equation}
where $\mathbf{F}$ is the time derivative of the state
that one stage of the integration returns. The vector $\mathbf{F}$ consists of
$\dot{\mathbf{u}}_b$, $\dot{\mathbf{u}}_t$, and $\dot{\mathbf{r}}$, and the
accelerations given by steps (i)--(iii) of Fig.~\ref{fig:dns}. The fluid has
no state of its own. Because the fluid is incompressible, the fluid pressure is fixed
at each instant by $\mathbf{w}$ through the linear solve of step (ii) of Fig.~\ref{fig:dns}
(Sec.~\ref{sec:incorporating}). The
pressure is thus computed within $\mathbf{F}$ but is not part of
$\mathbf{w}$.

With $\gamma = \gamma_0(x)$ and $f = 0$,
$\mathbf{F}(\mathbf{w}) \equiv \mathbf{F}(\mathbf{w}, 0)$ is linear in
$\mathbf{w}$ and equals $\mathsf{J}\mathbf{w}$ of
Eq.~(\ref{eq:linear_system}). The matrix $\mathsf{J}$ has
$(4N_x + 6) \times (4N_x + 6)$ entries.

The entries of $\mathsf{J}$ are not derived by hand but computed from the
discretized equations of motion. Let $\mathbf{e}_j$ be the state in which
only the $j$th component is 1. Any state is
$\mathbf{w} = \sum_j w_j \mathbf{e}_j$, and the linearity of $\mathbf{F}$
gives
\begin{equation}
\mathbf{F}(\mathbf{w}) = \sum_{j=1}^{N} w_j\, \mathbf{F}(\mathbf{e}_j),
\qquad N = 4N_x + 6,
\label{eq:superposition}
\end{equation}
so that the matrix
\begin{equation}
\mathsf{J} = \left[ \mathbf{F}(\mathbf{e}_1),\ \mathbf{F}(\mathbf{e}_2),\
\ldots,\ \mathbf{F}(\mathbf{e}_N) \right]
\label{eq:J_columns}
\end{equation}
satisfies Eq.~(\ref{eq:linear_system}) for every $\mathbf{w}$. Each column
is computed by one evaluation of the right-hand side, that is, by one
stage of the time integration without advancing in
time. This stage is shown in Fig.~\ref{fig:dns}. For random $\mathbf{w}$, $\mathsf{J}\mathbf{w}$
reproduces $\mathbf{F}(\mathbf{w})$ to machine precision.
All eigenvalues and eigenvectors of $\mathsf{J}$ ($8006 \times 8006$ for
$N_x = 2000$) are computed by the QR algorithm (LAPACK \texttt{dgeev}).

\section{The main line and the pump}
\label{app:pump}

This appendix gives the oscillation of the main line and $\mathsf{J}_0$
and $\mathsf{P}$ of Eq.~(\ref{eq:pair}). The model is written as
\begin{equation}
\dot{\mathbf{w}} = \mathsf{J}\mathbf{w} + \mathsf{B}\, \mathcal{N}, \qquad
\mathcal{N} = -\frac{\gamma_0}{u_{\mathrm{sat}}^2}\, u_c^2 \left( c_4 \dot{u}_c + k_4 u_c \right),
\label{eq:app_N}
\end{equation}
where $\mathcal{N}$ is the cubic part of the active force at each grid point
and $\mathsf{B}$ maps $\mathcal{N}$ to $\dot{\mathbf{w}}$.

On a single line, $u_c = A\cos\omega t$, the cubic term also produces the
frequency $3\omega$, but only the component of the cubic term at $\omega$ affects the line.
With $\cos^3\omega t = \tfrac{3}{4}\cos\omega t + \tfrac{1}{4}\cos 3\omega t$
and $\cos^2\omega t \sin\omega t = \tfrac{1}{4}\sin\omega t + \tfrac{1}{4}\sin 3\omega t$,
this component is
\begin{equation}
\mathcal{N}_\omega = -\gamma_0 S \left( c_4 \dot{u}_c + 3 k_4 u_c \right),
\qquad S = \frac{A^2}{4 u_{\mathrm{sat}}^2}.
\label{eq:df_single}
\end{equation}
The line thus sees the gains $\gamma_0(1 - S)$ in the damping and
$\gamma_0(1 - 3S)$ in the stiffness. This reduction is the self-saturation.

The oscillation of the main
line, $\mathbf{w} = \mathrm{Re}(\mathbf{W} e^{i\omega t})$, is found from
$i\omega \mathbf{W} = \mathsf{J}\mathbf{W} + \mathsf{B}\mathcal{N}_\omega$,
where $\mathcal{N}_\omega$ is the component of $\mathcal{N}$ at $\omega$
of Eq.~(\ref{eq:df_single}). As a phasor, $\mathcal{N}_\omega$ has the
entries $-\gamma_0 S (i\omega c_4 + 3k_4) U$, with $U$ the entries of
$\mathsf{C}_u\mathbf{W}$ of the next paragraph and $S = |U|^2/4u_{\mathrm{sat}}^2$. Since $S$ is not analytic in
$U$, the equation is split into its real and imaginary parts. The solution
is fixed in phase by setting the imaginary part of $\dot{u}_b$ to zero at
the node where $|\dot{u}_b|$ of the initial guess is largest. The resulting
$2N + 1$ real equations are solved for the real and imaginary parts of
$\mathbf{W}$ and for $\omega$ by Newton's method, starting from an
eigenvector of the mode of the main line.

Let the main line be $u_c = \mathrm{Re}(U e^{i\omega t})$ at each grid point, and let $\mathcal{S} = U^2/4u_{\mathrm{sat}}^2$, so that $|\mathcal{S}| = S$. A small change of $u_c$ and $\dot{u}_c$ changes $\mathcal{N}$ by the
derivatives $\partial\mathcal{N}/\partial u_c$ and
$\partial\mathcal{N}/\partial \dot{u}_c$, taken on the oscillation of the
main line. With $\dot u_c = \mathrm{Re}(i\omega U e^{i\omega t})$, these derivatives are
\begin{align}
\frac{\partial\mathcal{N}}{\partial \dot{u}_c}
&= -\gamma_0 c_4 \left( 2S + \mathcal{S} e^{2i\omega t} + \bar{\mathcal{S}} e^{-2i\omega t} \right), \nonumber\\
\frac{\partial\mathcal{N}}{\partial u_c}
&= -\gamma_0 \left[ 6S k_4 + (3k_4 + 2i\omega c_4)\, \mathcal{S} e^{2i\omega t} + (3k_4 - 2i\omega c_4)\, \bar{\mathcal{S}} e^{-2i\omega t} \right].
\label{eq:dN}
\end{align}
Let $\mathsf{C}_u$ and $\mathsf{C}_v$ pick $u_c$ and $\dot{u}_c$ at the grid points from $\mathbf{w}$, and let $S$ and $\mathcal{S}$ also denote the diagonal matrices of their values at the grid points. The matrix $S$ is real, and $\mathcal{S}$ is complex, with the conjugate $\bar{\mathcal{S}}$. The constant parts give
\begin{equation}
\mathsf{J}_0 = \mathsf{J} - \gamma_0 \mathsf{B} \left( 6 S k_4 \mathsf{C}_u + 2 S c_4 \mathsf{C}_v \right).
\label{eq:J0}
\end{equation}
The matrix $\mathsf{J}_0$ thus has the gains $\gamma_0(1 - 2S)$ in the damping and $\gamma_0(1 - 6S)$ in the stiffness. This lowering is the cross-saturation. The cross-saturation is thus twice the self-saturation of
Eq.~(\ref{eq:df_single}). The parts at $e^{2i\omega t}$ give the pump
\begin{equation}
\mathsf{P} = -\gamma_0 \mathsf{B} \left[ (3k_4 + 2i\omega c_4)\, \mathcal{S}\, \mathsf{C}_u + c_4 \mathcal{S}\, \mathsf{C}_v \right].
\label{eq:P}
\end{equation}
The parts at $e^{-2i\omega t}$ give $\bar{\mathsf{P}}$. These parts give Eq.~(\ref{eq:lin_main}). Substituting the complex solution
$\mathbf{c}_+ e^{(s + i\omega)t} + \mathbf{c}_- e^{(s - i\omega)t}$
into Eq.~(\ref{eq:lin_main}) and collecting the terms at
$e^{(s + i\omega)t}$ and $e^{(s - i\omega)t}$ gives Eq.~(\ref{eq:pair}).
Terms at $\pm 3\omega$ are neglected.

The eigenvalues of Eq.~(\ref{eq:pair}), a matrix of size
$2(4N_x + 6)$, are found by shift-and-invert with shifts every 200~Hz, from 5.4 to
7.2~kHz.
Since $s$ and $\bar s$ describe the same pair with the two lines exchanged,
the shifts cover both sides of the main line. These eigenvalues are the
Floquet exponents of the oscillation of the main line, truncated at
$\pm\omega$.

\let\thinmuskip\JASAsavedthinmuskip
\let\tmspace\JASAsavedtmspace
\bibliography{refs}

@article{LiuNeely2010,
  author  = {Liu, Yi-Wen and Neely, Stephen T.},
  title   = {Distortion product emissions from a cochlear model with nonlinear mechanoelectrical transduction in outer hair cells},
  journal = {J. Acoust. Soc. Am.},
  volume  = {127},
  number  = {4},
  pages   = {2420--2432},
  year    = {2010}
}

@article{Murakami2019,
  author  = {Murakami, Yasuki},
  title   = {Efficiency limit of nonuniform grid setting in two-dimensional cochlear model},
  journal = {Acoust. Sci. Tech.},
  volume  = {40},
  number  = {5},
  pages   = {336--343},
  year    = {2019},
  doi     = {10.1250/ast.40.336}
}

@article{Murakami2021,
  author  = {Murakami, Yasuki},
  title   = {Fast time-domain solution of a nonlinear three-dimensional cochlear model using the fast {F}ourier transform},
  journal = {J. Acoust. Soc. Am.},
  volume  = {150},
  number  = {4},
  pages   = {2589--2599},
  year    = {2021}
}

@article{Sisto2019,
  author  = {Sisto, Renata and Shera, Christopher A. and Alto{\`e}, Alessandro and Moleti, Arturo},
  title   = {Constraints imposed by zero-crossing invariance on cochlear models with two mechanical degrees of freedom},
  journal = {J. Acoust. Soc. Am.},
  volume  = {146},
  number  = {3},
  pages   = {1685--1695},
  year    = {2019}
}

@article{Sisto2021,
  author  = {Sisto, Renata and Belardinelli, Daniele and Moleti, Arturo},
  title   = {Fluid focusing and viscosity allow high gain and stability of the cochlear response},
  journal = {J. Acoust. Soc. Am.},
  volume  = {150},
  number  = {6},
  pages   = {4283--4296},
  year    = {2021}
}

@article{Shera2003,
  author  = {Shera, Christopher A.},
  title   = {Mammalian spontaneous otoacoustic emissions are amplitude-stabilized cochlear standing waves},
  journal = {J. Acoust. Soc. Am.},
  volume  = {114},
  pages   = {244--262},
  year    = {2003}
}

@book{Lamb1932,
  author    = {Lamb, Horace},
  title     = {Hydrodynamics},
  edition   = {6th},
  publisher = {Cambridge University Press},
  address   = {Cambridge},
  year      = {1932}
}

@article{Stokes1851,
  author  = {Stokes, G. G.},
  title   = {On the effect of the internal friction of fluids on the motion of pendulums},
  journal = {Trans. Cambridge Philos. Soc.},
  volume  = {9},
  number  = {Part II},
  pages   = {8--106},
  year    = {1851}
}

@article{ZweigShera1995,
  author  = {Zweig, George and Shera, Christopher A.},
  title   = {The origin of periodicity in the spectrum of evoked otoacoustic emissions},
  journal = {J. Acoust. Soc. Am.},
  volume  = {98},
  number  = {4},
  pages   = {2018--2047},
  year    = {1995},
  doi     = {10.1121/1.413320}
}

@article{Talmadge1998,
  author  = {Talmadge, Carrick L. and Tubis, Arnold and Long, Glenis R. and Piskorski, Pawel},
  title   = {Modeling otoacoustic emission and hearing threshold fine structures},
  journal = {J. Acoust. Soc. Am.},
  volume  = {104},
  number  = {3},
  pages   = {1517--1543},
  year    = {1998}
}

@article{Shera2022,
  author  = {Shera, Christopher A.},
  title   = {Whistling while it works: Spontaneous otoacoustic emissions and the cochlear amplifier},
  journal = {J. Assoc. Res. Otolaryngol.},
  volume  = {23},
  pages   = {17--25},
  year    = {2022},
  doi     = {10.1007/s10162-021-00829-9}
}

@article{Elliott2007,
  author  = {Elliott, Stephen J. and Ku, Emery M. and Lineton, Ben},
  title   = {A state space model for cochlear mechanics},
  journal = {J. Acoust. Soc. Am.},
  volume  = {122},
  number  = {5},
  pages   = {2759--2771},
  year    = {2007},
  doi     = {10.1121/1.2783125}
}

@article{Ku2008,
  author  = {Ku, Emery M. and Elliott, Stephen J. and Lineton, Ben},
  title   = {Statistics of instabilities in a state space model of the human cochlea},
  journal = {J. Acoust. Soc. Am.},
  volume  = {124},
  number  = {2},
  pages   = {1068--1079},
  year    = {2008},
  doi     = {10.1121/1.2939133}
}

@article{Ku2009,
  author  = {Ku, Emery M. and Elliott, Stephen J. and Lineton, Ben},
  title   = {Limit cycle oscillations in a nonlinear state space model of the human cochlea},
  journal = {J. Acoust. Soc. Am.},
  volume  = {126},
  number  = {2},
  pages   = {739--750},
  year    = {2009},
  doi     = {10.1121/1.3158861}
}

@inproceedings{Meaud2018,
  author    = {Meaud, Julien and Bowling, Thomas and Lemons, Charlsie},
  title     = {Computational modeling of spontaneous otoacoustic emissions by the mammalian cochlea},
  booktitle = {Proceedings of the ASME 2018 Dynamic Systems and Control Conference},
  volume    = {1},
  pages     = {V001T11A001},
  year      = {2018},
  doi       = {10.1115/DSCC2018-9044}
}

@article{Bowling2019,
  author  = {Bowling, Thomas and Lemons, Charlsie and Meaud, Julien},
  title   = {Reducing tectorial membrane viscoelasticity enhances spontaneous otoacoustic emissions and compromises the detection of low level sound},
  journal = {Sci. Rep.},
  volume  = {9},
  pages   = {7494},
  year    = {2019},
  doi     = {10.1038/s41598-019-43970-5}
}

@article{Nobili2003,
  author  = {Nobili, Renato and Vete{\v{s}}n{\'i}k, Ale{\v{s}} and Turicchia, Lorenzo and Mammano, Fabio},
  title   = {Otoacoustic emissions from residual oscillations of the cochlear basilar membrane in a human ear model},
  journal = {J. Assoc. Res. Otolaryngol.},
  volume  = {4},
  pages   = {478--494},
  year    = {2003},
  doi     = {10.1007/s10162-002-3055-1}
}

@article{deBoerNuttall2006,
  author  = {de Boer, Egbert and Nuttall, Alfred L.},
  title   = {Spontaneous basilar-membrane oscillation ({SBMO}) and coherent reflection},
  journal = {J. Assoc. Res. Otolaryngol.},
  volume  = {7},
  number  = {1},
  pages   = {26--37},
  year    = {2006},
  doi     = {10.1007/s10162-005-0020-9}
}

@article{Nuttall2004,
  author  = {Nuttall, A. L. and Grosh, K. and Zheng, J. and de Boer, E. and Zou, Y. and Ren, T.},
  title   = {Spontaneous basilar membrane oscillation and otoacoustic emission at 15 {kHz} in a guinea pig},
  journal = {J. Assoc. Res. Otolaryngol.},
  volume  = {5},
  pages   = {337--348},
  year    = {2004},
  doi     = {10.1007/s10162-004-4045-2}
}

@article{Vencovsky2020,
  author  = {Vencovsk{\'y}, V{\'a}clav and Vete{\v{s}}n{\'i}k, Ale{\v{s}} and Gummer, Anthony W.},
  title   = {Nonlinear reflection as a cause of the short-latency component in stimulus-frequency otoacoustic emissions simulated by the methods of compression and suppression},
  journal = {J. Acoust. Soc. Am.},
  volume  = {147},
  number  = {6},
  pages   = {3992--4008},
  year    = {2020},
  doi     = {10.1121/10.0001394}
}

@article{vanDijkWit1998,
  author  = {van Dijk, Pim and Wit, Hero P.},
  title   = {Synchronization of cubic distortion spontaneous otoacoustic emissions},
  journal = {J. Acoust. Soc. Am.},
  volume  = {104},
  number  = {1},
  pages   = {591--594},
  year    = {1998},
  doi     = {10.1121/1.423287}
}

@article{Murakami2018,
  author  = {Murakami, Yasuki},
  title   = {Influence of discretization error on instability of cochlear model},
  journal = {J. Acoust. Soc. Am.},
  volume  = {144},
  number  = {6},
  pages   = {3593--3602},
  year    = {2018},
  doi     = {10.1121/1.5083836}
}

@article{MurakamiFuji2022,
  author  = {Murakami, Yasuki and Fuji, Takumi},
  title   = {Difference between frequency and suppression tuning curves in a two-dimensional cochlear model},
  journal = {JASA Express Lett.},
  volume  = {2},
  pages   = {094402},
  year    = {2022},
  doi     = {10.1121/10.0013998}
}

@article{Charaziak2014,
  author  = {Charaziak, Karolina K. and Siegel, Jonathan H.},
  title   = {Estimating cochlear frequency selectivity with stimulus-frequency otoacoustic emissions in chinchillas},
  journal = {J. Assoc. Res. Otolaryngol.},
  volume  = {15},
  pages   = {883--896},
  year    = {2014},
  doi     = {10.1007/s10162-014-0487-3}
}

@article{Kemp1978, title={Stimulated acoustic emissions from within the human auditory system}, volume={64}, DOI={10.1121/1.382104}, number={5}, journal={The Journal of the Acoustical Society of America}, author={Kemp, D. T.}, year={1978}, pages={1386–1391} }

@article{SheraGuinan2003, title={Stimulus-frequency-emission group delay: A test of coherent reflection filtering and a window on cochlear tuning}, volume={113}, DOI={10.1121/1.1557211}, number={5}, journal={The Journal of the Acoustical Society of America}, author={Shera, Christopher A. and Guinan, John J.}, year={2003}, pages={2762–2772} }

@article{Camalet2000, title={Auditory sensitivity provided by self-tuned critical oscillations of hair cells}, volume={97}, DOI={10.1073/pnas.97.7.3183}, number={7}, journal={Proceedings of the National Academy of Sciences}, author={Camalet, Sébastien and Duke, Thomas and Jülicher, Frank and Prost, Jacques}, year={2000}, pages={3183–3188} }

@article{Hudspeth2010, title={A Critique of the Critical Cochlea: Hopf—a Bifurcation—Is Better Than None}, volume={104}, DOI={10.1152/jn.00437.2010}, number={3}, journal={Journal of Neurophysiology}, author={Hudspeth, A. J. and Jülicher, Frank and Martin, Pascal}, year={2010}, pages={1219–1229} }

@article{WitVanDijk2012, title={Are human spontaneous otoacoustic emissions generated by a chain of coupled nonlinear oscillators?}, volume={132}, DOI={10.1121/1.4730886}, number={2}, journal={The Journal of the Acoustical Society of America}, author={Wit, Hero P. and van Dijk, Pim}, year={2012}, pages={918–926} }

@article{WitBell2024, title={Something in Our Ears Is Oscillating, but What? A Modeller’s View of Efforts to Model Spontaneous Emissions}, volume={25}, DOI={10.1007/s10162-024-00940-7}, number={4}, journal={Journal of the Association for Research in Otolaryngology}, author={Wit, Hero P. and Bell, Andrew}, year={2024}, pages={313–328} }

@article{Wit2026, title={A Time-Domain Model for the Generation of Spontaneous Otoacoustic Emissions in Resonant Cavities}, volume={27}, DOI={10.1007/s10162-026-01063-x}, number={5}, journal={Journal of the Association for Research in Otolaryngology}, author={Wit, Hero P.}, year={2026}, pages={785–804} }

@article{Diependaal1987, title={Numerical methods for solving one-dimensional cochlear models in the time domain}, volume={82}, DOI={10.1121/1.395157}, number={5}, journal={The Journal of the Acoustical Society of America}, author={Diependaal, Rob J. and Duifhuis, H. and Hoogstraten, H. W. and Viergever, Max A.}, year={1987}, pages={1655–1666} }

@article{Verhulst2012, title={Nonlinear time-domain cochlear model for transient stimulation and human otoacoustic emission}, volume={132}, DOI={10.1121/1.4763989}, number={6}, journal={The Journal of the Acoustical Society of America}, author={Verhulst, Sarah and Dau, Torsten and Shera, Christopher A.}, year={2012}, pages={3842–3848} }

@article{AllenSondhi1979, title={Cochlear macromechanics: Time domain solutions}, volume={66}, DOI={10.1121/1.383064}, number={1}, journal={The Journal of the Acoustical Society of America}, author={Allen, J. B. and Sondhi, M. M.}, year={1979}, pages={123–132} }

@article{Cheatham2018, title={Spontaneous otoacoustic emissions in {Tecta}$^{Y1870C/+}$ mice reflect changes in cochlear amplification and how it is controlled by the tectorial membrane}, volume={5}, DOI={10.1523/eneuro.0314-18.2018}, number={6}, journal={eNeuro}, author={Cheatham, Mary Ann and Zhou, Yingjie and Goodyear, Richard J. and Dallos, Peter and Richardson, Guy P.}, year={2018}, pages={ENEURO.0314–18.2018} }

@article{Shera1999, title={Evoked otoacoustic emissions arise by two fundamentally different mechanisms: A taxonomy for mammalian {OAEs}}, volume={105}, DOI={10.1121/1.426948}, number={2}, journal={The Journal of the Acoustical Society of America}, author={Shera, Christopher A. and Guinan, John J.}, year={1999}, pages={782–798} }

@article{vanDijkWit1990, title={Amplitude and frequency fluctuations of spontaneous otoacoustic emissions}, volume={88}, DOI={10.1121/1.400199}, number={4}, journal={The Journal of the Acoustical Society of America}, author={van Dijk, Pim and Wit, Hero P.}, year={1990}, pages={1779–1793} }

@article{Burns2009, title={Long-term stability of spontaneous otoacoustic emissions}, volume={125}, DOI={10.1121/1.3097768}, number={5}, journal={The Journal of the Acoustical Society of America}, author={Burns, Edward M.}, year={2009}, pages={3166–3176} }

@article{Bergevin2025, title={Auditory cellular cooperativity probed via spontaneous otoacoustic emissions}, volume={124}, DOI={10.1016/j.bpj.2025.02.023}, number={8}, journal={Biophysical Journal}, author={Bergevin, Christopher and Whiley, Rebecca E. and Wit, Hero and Manley, Geoffrey A. and van Dijk, Pim}, year={2025}, pages={1208–1225} }

@article{Gelfand2010, title={Interactions between Hair Cells Shape Spontaneous Otoacoustic Emissions in a Model of the Tokay Gecko’s Cochlea}, volume={5}, DOI={10.1371/journal.pone.0011116}, number={6}, journal={PLoS ONE}, author={Gelfand, Michael and Piro, Oreste and Magnasco, Marcelo O. and Hudspeth, A. J.}, editor={Brezina, Vladimir}, year={2010}, pages={e11116} }

@article{RoblesRuggero2001, title={Mechanics of the Mammalian Cochlea}, volume={81}, DOI={10.1152/physrev.2001.81.3.1305}, number={3}, journal={Physiological Reviews}, author={Robles, Luis and Ruggero, Mario A.}, year={2001}, pages={1305–1352} }

@article{Fruth2014, title={An active oscillator model describes the statistics of spontaneous otoacoustic emissions}, volume={107}, DOI={10.1016/j.bpj.2014.06.047}, number={4}, journal={Biophysical Journal}, author={Fruth, Florian and J{\"u}licher, Frank and Lindner, Benjamin}, year={2014}, pages={815--824} }

@article{Lamb1964,
  author  = {Lamb, Jr., Willis E.},
  title   = {Theory of an optical maser},
  journal = {Phys. Rev.},
  volume  = {134},
  pages   = {A1429--A1450},
  year    = {1964},
  doi     = {10.1103/PhysRev.134.A1429}
}

@article{Kippenberg2011,
  author  = {Kippenberg, Tobias J. and Holzwarth, Ronald and Diddams, Scott A.},
  title   = {Microresonator-based optical frequency combs},
  journal = {Science},
  volume  = {332},
  pages   = {555--559},
  year    = {2011},
  doi     = {10.1126/science.1193968}
}

@article{Emelianova2014,
  author  = {Emelianova, Yulia P. and Emelyanov, Valeriy V. and Ryskin, Nikita M.},
  title   = {Synchronization of two coupled multimode oscillators with time-delayed feedback},
  journal = {Commun. Nonlinear Sci. Numer. Simul.},
  volume  = {19},
  pages   = {3778--3791},
  year    = {2014},
  doi     = {10.1016/j.cnsns.2014.03.031}
}

@article{Agrawal1988,
  author  = {Agrawal, Govind P.},
  title   = {Population pulsations and nondegenerate four-wave mixing in semiconductor lasers and amplifiers},
  journal = {J. Opt. Soc. Am. B},
  volume  = {5},
  pages   = {147--159},
  year    = {1988},
  doi     = {10.1364/JOSAB.5.000147}
}

@article{Dewey2025, title={Sources of Microstructure in Mammalian Cochlear Responses}, volume={26}, DOI={10.1007/s10162-025-00974-5}, number={1}, journal={Journal of the Association for Research in Otolaryngology}, author={Dewey, James B.}, year={2025}, pages={1--15} }

@article{BianWatts2008, title={Effects of low-frequency biasing on spontaneous otoacoustic emissions: {A}mplitude modulation}, volume={123}, DOI={10.1121/1.2821983}, number={2}, journal={The Journal of the Acoustical Society of America}, author={Bian, Lin and Watts, Kelly L.}, year={2008}, pages={887--898} }

@article{Khurgin2014, title={Coherent frequency combs produced by self frequency modulation in quantum cascade lasers}, volume={104}, DOI={10.1063/1.4866868}, number={8}, journal={Applied Physics Letters}, author={Khurgin, J. B. and Dikmelik, Y. and Hugi, A. and Faist, J.}, year={2014}, pages={081118} }

@article{Hugi2012, title={Mid-infrared frequency comb based on a quantum cascade laser}, volume={492}, DOI={10.1038/nature11620}, number={7428}, journal={Nature}, author={Hugi, Andreas and Villares, Gustavo and Blaser, St{\'e}phane and Liu, H. C. and Faist, J{\'e}r{\^o}me}, year={2012}, pages={229--233} }

@article{Rhode2007, title={Mutual suppression in the 6 {kHz} region of sensitive chinchilla cochleae}, volume={121}, DOI={10.1121/1.2718398}, number={5}, journal={The Journal of the Acoustical Society of America}, author={Rhode, William S.}, year={2007}, pages={2805--2818} }

\end{document}